\documentclass[12pt]{article}
\usepackage[a4paper, hmargin=2.5cm, vmargin=2.5cm]{geometry}
\usepackage[authoryear]{natbib} \bibpunct{(}{)}{,}{a}{,}{,}
\usepackage{amsmath}
\usepackage{graphicx}
\usepackage{mathrsfs}
\usepackage{amssymb}
\usepackage{verbatim}
\usepackage{comment}

\usepackage{url}
\usepackage{booktabs}
\usepackage{color, colortbl}
\usepackage{lscape}
\usepackage{bm}
\definecolor{Gray}{gray}{0.9}
\usepackage{graphicx}
\usepackage{float}

\usepackage{color}

\newcommand{\cD}{\mathscr{D}}

\newcommand{\bA}{\mbox{\boldmath $A$}}
\newcommand{\bB}{\mbox{\boldmath $B$}}

\newcommand{\bO}{\bf O}
\newcommand{\om}{\overline{m}}
\newcommand{\bW}{\mbox{\boldmath $W$}}

\newcommand{\bH}{\mbox{\boldmath $H$}}
\newcommand{\bI}{\mbox{\boldmath $I$}}
\newcommand{\bJ}{\mbox{\boldmath $J$}}

\newcommand{\bV}{\mbox{\boldmath $V$}}
\newcommand{\bY}{\mbox{\boldmath $Y$}}
\newcommand{\by}{\mbox{\boldmath $y$}}
\newcommand{\bX}{\mbox{\boldmath $X$}}
\newcommand{\bx}{\mbox{\boldmath $x$}}

\newcommand{\bzero}{\mbox{\boldmath $0$}}
\newcommand{\bmu}{\mbox{\boldmath $\mu$}}
\newcommand{\bbeta}{\mbox{\boldmath $\beta$}}

\newcommand{\bomega}{\mbox{\boldmath $\omega$}}
\newcommand{\bLambda}{\mbox{\boldmath $\Lambda$}}

\newcommand{\hR}{\hat{\mbox{$R$}}}
\newcommand{\hbeta}{\hat{\mbox{$\beta$}}}

\newcommand{\bSigma}{\mbox{\boldmath $\Sigma$}}
\newcommand{\bPsi}{\mbox{\boldmath $\Psi$}}

\newcommand{\bxi}{\mbox{\boldmath $\xi$}}
\newcommand{\btheta}{\mbox{\boldmath $\theta$}}

\newcommand{\hbbeta}{\hat{\mbox{\boldmath $\bbeta$}}}

\newcommand{\vs}{\vspace{0.2cm}}
\newcommand{\mat}[1]{\bm{#1}}

\newcommand{\vect}[1]{\bm{#1}}

\newcommand{\T}{T}

\begin{document}

\noindent
An Apparent Paradox: A Classifier Trained from a Partially Classified Sample 
May Have Smaller Expected Error Rate Than That If the 
Sample Were Completely Classified
\vs
\vs

\noindent
Daniel Ahfock* and Geoffrey J. McLachlan \newline
School of Mathematics and Physics, University of Queensland, Brisbane \newline
\vspace*{-0.5cm} \ \\
*email:\url{d.ahfock@uq.edu.au}

\vs\vs\vs
\noindent
{\bf Abstract:}
\vs

\noindent
There has been increasing interest in using semi-supervised learning 
to form a classifier. As is well known, the (Fisher) information 
in an unclassified feature with unknown class label is less (considerably 
less for weakly separated classes) than that of a classified feature 
which has known class label. Hence assuming that the labels of the 
unclassified features are randomly missing or their missing-label mechanism 
is simply ignored, the expected error rate of a classifier formed 
from a partially classified sample is greater than that if the sample 
were completely classified. We propose to treat the labels of the 
unclassified features as missing data and to introduce a framework 
for their missingness in situations where these labels are not 
randomly missing. An examination of several partially classified data sets 
in the literature suggests that the unclassified features are not occurring 
at random but rather tend to be concentrated in regions of 
relatively high entropy in the feature space. Here in the context 
of two normal classes with a common covariance matrix 
we consider the situation where the missingness of the labels 
of the unclassified features can be modelled by a logistic model 
in which the probability of a missing label for a feature depends 
on its entropy. Rather paradoxically, we show that the classifier 
so formed from the partially classified sample may have smaller 
expected error rate that that if the sample were completely classified.
\vs
\vs

\section{Introduction}
We consider the problem of forming a classifier from training data that 
are not completely classified. That is, the feature vectors $\by_j$ in the
training sample have all been observed but their class labels are missing 
for some of them and so the training data constitute a partially
classified sample denoted here by $\bx_{\rm PC}$.
This problem goes back at least to the mid-seventies
\citep{mclachlan_1975_iterative}, and it received a boost shortly afterwards 
with the advent of the EM algorithm
\citep{dempster_1977_maximum} which could be applied to carry out
maximum likelihood (ML) estimation for a partially classified sample.
These days increasing attention is being given to the formation of
classifiers on the basis of a partially classified sample
(or semi-supervised learning (SSL) as it is referred to in the machine
learning literature), particularly
in situations where unclassified data are available more freely or
more cheaply or both than classified data.
Moreover, in some instances in the field of medical diagnosis,
a definitive classification can only be made via an invasive procedure that
may not be ethical to apply unless there is a high degree of confidence
that the patient has the disease for which screening is being performed.
There is now a wide literature on SSL techniques 
(for example, \cite{grandvalet_2005_semi} and \cite{berthelot_2019_mixmatch}),
which are too numerous to discuss here.

In SSL, it is usually assumed that the labels 
of the unclassified features are randomly missing or the
missing-label mechanism is simply ignored.  We propose a 
joint modelling framework that introduces a missing-label mechanism 
for the missing-label indicators which are treated as random variables. 
Our examination of a number of real datasets shows that 
the pattern of missing labels is typically related to the 
difficulty of classification, which can be quantified by 
the Shannon entropy.  This relationship can be captured using 
a logistic selection model. Full likelihood inference that includes 
the missing-label mechanism can improve the efficiency of 
parameter estimation and increase classification accuracy to the extent 
where it can be greater than if the sample were completely classified.
\vs

More specifically, 
we let $m_j$ be the missing-label indicator being equal to 1 if 
the $j$th feature vector in the training sample is
unclassified; that is, its class label is missing.
In the case of a partially classified training sample
$\bx_{\rm PC}$ in the context of the 
two-class normal discrimination problem, 
\citet{oneill_1978_normal} showed
that the information about the vector $\bbeta$ of discriminant 
function coefficients using the likelihood
that ignores the mechanism for the missing labels can be decomposed as
\begin{equation}
\bI_{\rm PC}^{(\rm ig)}(\bbeta)=\bI_{\rm CC}(\bbeta) 
-\om\bI_{\rm CC}^{(\rm lr)}(\bbeta),
\label{eq:g1}
\end{equation}
where $\bI_{\rm CC}(\bbeta)$ is the information about $\bbeta$ in a completely
classified sample $\bx_{\rm CC}, \bI_{\rm CC}^{\rm(clr)}(\bbeta)$
is the information about $\bbeta$ under the logistic regression model 
for the distribution of the class labels given the features in $\bx_{\rm CC}$, 
and $\om=\sum_{j=1}^n m_j/n$
is the proportion of unclassified features in the partially classified 
sample $\bx_{\rm PC}$.
It can be seen from (\ref{eq:g1}) that the loss of information due to
the sample being partially classified is equal to
$\om\bI_{\rm CC}^{(\rm lr)}(\bbeta)$.
The consequent decrease in the efficiency in estimating the Bayes' rule
can be considerable as illustrated in Table \ref{tab:mcar} in Section 5.
\vs

With our proposed approach, we introduce the random variable
$M_j$ corresponding to the realized value $m_j$ for the 
missing-label indicator for the feature vector $\by_j$ and model its distribution
to depend on an entropy-based measure.
We then consider the estimation of $\bbeta$ from the partially
classified sample $\bx_{\rm PC}$ on the basis of the so-called full likelihood 
$L_{\rm PC}^{(\rm full)}(\btheta)$ whose logarithm is augmented by the addition
of the log likelihood for $\bbeta$ formed under the proposed logistic model for the
missing-label indicator random variable $M_j$.
We then show that the information about $\bbeta$ for the full likelihood 
formed from the partially classified sample $\bx_{\rm PC}$ is given by
\begin{equation}
\bI_{\rm PC}^{(\rm full)}(\bbeta)= \bI_{\rm CC}(\bbeta) 
-\gamma\bI_{\rm CC}^{(\rm clr)}(\bbeta) + 
\bI_{\rm PC}^{(\rm miss)}(\bbeta),
\label{eq:g2}
\end{equation}
where
$\bI_{\rm CC}^{(\rm clr)}(\bbeta)$ is the conditional information about
$\bbeta$ under the logistic regression model fitted to the class
labels in $\bx_{\rm CC},
\bI_{\rm PC}^{(\rm miss)}(\bbeta)$ is the information about
$\bbeta$ in the missing-label indicators $m_j$,
and $\gamma$ is the expected proportion of missing class
labels in the partially classified sample.
It can be seen from (\ref{eq:g2}) that if 
$$\bI_{\rm PC}^{(\rm miss)}(\bbeta) > 
\gamma\bI_{\rm CC}^{(\rm clr)}(\bbeta),$$
then there is actually an increase in the information about $\bbeta$ 
in the partially classified sample over
the information $\bI_{\rm CC}(\bbeta)$ about $\bbeta$
in the completely classified sample. Here, the inequality in the above equation is used in the sense that the left-hand side of the equation, minus the right, is positive definite.
Following on from \citet{ahfock_2019_talk},
we shall show that 
under certain conditions on the distribution of the missing labels that the
consequent reduction in the asymptotic expected error rate of the 
Bayes' rule learnt
using the partially classified sample is lower
than that of the Bayes' rule learnt using a completely 
classified sample.
Some Monte Carlo simulations are to be given to support 
the asymptotic theory.

\section{Two-Class Normal Discrimination}

In discriminant analysis, the aim is to assign an unclassified entity 
with $p$-dimensional feature vector $\by$ to one of a number of $g$ classes 
$C_1,\,\ldots,\,C_g$. It is assumed that the random vector $\bY$
corresponding to $\by$ has density $f_i(\by;\bomega_i)$ in $C_i$, 
specified up to an unknown vector of parameters $\bomega_i\,
(i = 1\,\ldots,\,g)$.
We consider here the case of $g=2$  classes for which
$f_i(\by;\bomega_i)$ denotes the multivariate normal density
with mean $\bmu_i$ and covariance matrix $\bSigma\,(i=1,2)$.
We let $\btheta$ be the vector containing the mixing proportion $\pi_1$,
the $2p$ elements of the means $\bmu_1$\ and $\bmu_2$, and the
${\textstyle\frac{1}{2}}p(p+1)$ elements of the common class-covariance matrix
$\bSigma$ known {\it a priori} to be distinct.
\vs

We let $R(\by;\btheta)$ denote the Bayes' (optimal) rule of allocation, 
where $R(\by; \btheta) = h$, that is, $\by$ is allocated to $C_h$, if
$$ h = \arg \max_i \tau_i(\by;\bbeta),$$
where 
\begin{eqnarray}
\tau_1(\by;\bbeta)&=& 1- \tau_2(\by;\bbeta)
\nonumber\\
&=& \pi_1 f_1(\by;\bomega_1)/f_y(\by;\btheta)
\nonumber\\
&=& \exp(\bbeta_0+\bbeta_1^T\by)/
\{1+ \exp(\bbeta_0+\bbeta_1^T\by)\}
\label{eq:g3}
\end{eqnarray}
is the posterior probability that $\by$ belongs to $C_1$ given $\bY= \by$; 
see, for example, \citet[Chapter 1]{mclachlan_1992_discriminant}.
Here $f_y(\by;\btheta)=\sum_{i=1}^2\pi_i f_i(\by;\bomega_i)$ is the marginal
(mixture) density of $\bY$
and $\bbeta=(\beta_0,\bbeta_1^T)^T$ is the vector of discriminant 
function coefficients, where
\begin{eqnarray*}
\beta_0 &=& - {\textstyle\frac{1}{2}}(\bmu_1+\bmu_2)^T\bSigma^{-1}(\bmu_1-\bmu_2),\\
\bbeta_1 &=& \bSigma^{-1} (\bmu_1-\bmu_2).
\end{eqnarray*}

It can be seen from (\ref{eq:g3}) that the Bayes' rule reduces in this
case of $g=2$ normal classes with a common covariance matrix to depending 
only on $\bbeta$ with $R(\by;\bbeta)$
being equal to 1 or 2, according as the discriminant function
$$d(\by;\bbeta)= \beta_0 + \bbeta^T\by$$
is greater or less than zero. 
\vs

We henceforth adopt the canonical form
\begin{equation}
\bmu_1=-\bmu_2=({\textstyle\frac{1}{2}}\Delta,\,0,\dots,\,0)^T,
\quad \bSigma=\bI_p,
\label{eq:g2b}
\end{equation}
where
$\Delta^2=(\bmu_1-\bmu_2)^T\bSigma^{-1}(\bmu_1-\bmu_2)$
is the Mahalanobis squared distance between the two classes
and $\bI_p$ is the $p\times p$ identity matrix.
\vs

In practice, $\bbeta$ has to be estimated from available training data. 
We let 
$\bx_{\rm CC}=(\bx_1^T,\,\ldots,\,\bx_n^T)^T$
denote $n$ independent realizations of $\bX=(\bY^T, Z)^T$ 
as the completely classified training data, 
where $Z$ denotes the
class membership of $\bY$, being equal to 1 if $\bY$ 
belongs to $C_1$, and zero otherwise.
We let $m_j$ be the missing-label indicator being equal to 1 if $z_j$
is missing and zero if it is available $(j=1,\,\ldots,\,n)$.
Accordingly, the unclassified sample $\bx_{\rm PC}$ is given by those
members $\bx_j$ in $\bx_{\rm CC}$ for which $m_j=0$ and only the feature
vectors $\by_j$ without their class labels $z_j$
for those members in $\bx_{\rm CC}$
for which $m_j=1$.
\vs
It should be noted that in our notation to denote the
various information matrices $\bI(\cdot)$  about a parameter,
we only display that parameter in the argument of $\bI(\cdot)$,
although $\bI$ may depend also
on other parameters, including those in the distribution adopted
for the missing-label indicators.
\vs 

With our proposed approach to exploiting the potential information in
the missing-label indicators $m_j$,
we introduce the random variable
$M_j$ corresponding to the realized value $m_j$ for the 
missing-class label for the feature vector $\by_j$ and model its distribution
to depend on an entropy-based measure.

\section{Mechanism for Missing Class Labels}

In many applications, the class labels $z_j$ are often assigned
by domain experts, who may not be able
to make a confident classification for every feature.
As a motivating example for our approach
to the formulation of a model for the distribution of the missing-label 
indicator $M_j$, we present Figure 1,
which shows a manually classified flow
cytometry dataset from \cite{aghaeepour2013critical}.
Black squares correspond to unclassified features, 
and the majority of the unclassified features appear
to be located near class boundaries.
Plots of other such datasets may be found in \citet{ahfock_2019_missing}.
\vs

\begin{figure}
    \centering
    \includegraphics[width=0.5\textwidth]{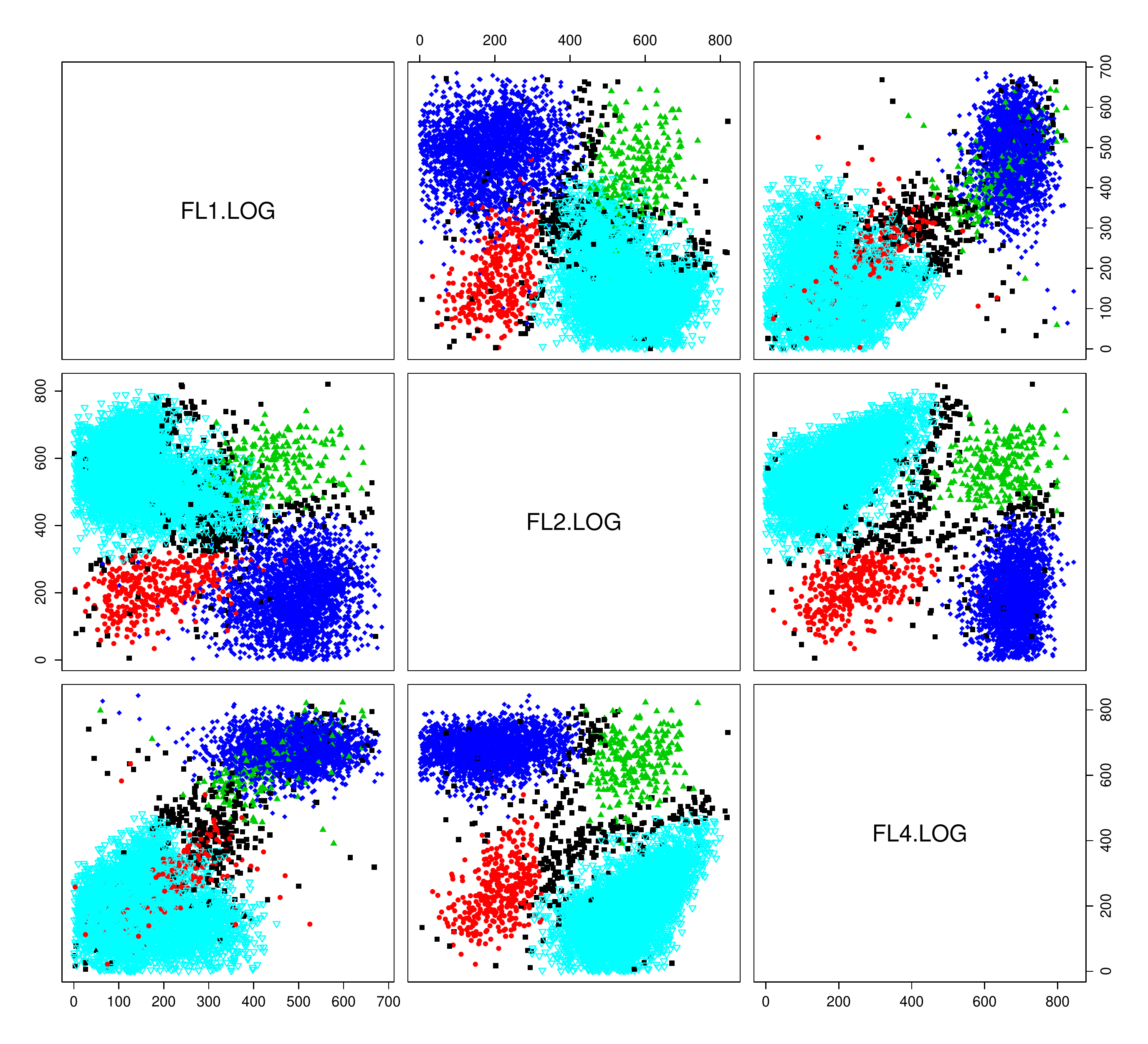}
    \caption{Flow cytometry dataset. Black squares correspond to unclassified observations.}
    \label{fig:cytometry}
\end{figure}

A standard approach in semi-supervised learning is to 
ignore the underlying cause in forming the likelihood
from the partially classified dataset. 
We shall denote this likelihood by $L_{\rm PC}^{(\rm ig)}(\btheta)$ with logarithm given by
\begin{equation}
\log L_{\rm PC}^{(\rm ig)}(\btheta)=
\sum_{j=1}^n(1-m_j)\sum_{i=1}^2 z_{ij}\log\{\pi_i f_i(\by_j;\bomega_i)\}
+
\sum_{j=1}^n m_j \log f_y(\by_j;\btheta),
\label{eq:g4}
\end{equation}
where $z_{1j}= 1-z_{2j} =z_j\,(j=1,\,\ldots,\,n).$
\vs

Note that the log of the likelihood $L_{\rm CC}(\btheta)$ for the
completely classified sample $\bx_{\rm CC}$ is given by (\ref{eq:g4})
with all $m_j=0$.
\vs

The missingness of class labels 
can be ignored in forming the likelihood function for $\btheta$
in the case of missing completely at random (MCAR) 
and for the less restrictive situation of missing at random (MAR).
However, in the latter situation, the (Fisher) information
will be affected by ignoring the missingness 
\citep{mclachlan1989mixture}.
\vs

If classification difficulty is a cause of the missing labels, 
the use of $L_{\rm PC}^{(\rm ig)}(\btheta)$
may be suboptimal.
In such circumstances,
the unlabelled features are likely to lie near class boundaries, and
then the pattern of missing labels
carries extra information for the estimation of $\btheta$ 
that is not reflected in (\ref{eq:g4}). 
The missing-data framework pioneered by \cite{rubin_1976}
is useful to exploit
the potential information in the missing-label pattern 
in the situation of a partially classified training sample
$\bx_{\rm PC}.$ 
We introduce the missing-label indicator random variable $M_j$ 
with realized value $m_j\,(j=1,\,\ldots,\,n)$.
An important measure of classification difficulty 
is the Shannon entropy of the posterior class probabilities.
Let $e_j$ denote the entropy for $\by_j$,
\begin{equation}
e_j=-\sum_{i=1}^2 \tau_i(\by_j;\bbeta)\log \tau_i(\by_j;\bbeta).
\label{eq:g5}
\end{equation}

Under our proposed missing-label model, we have that
\begin{eqnarray}
{\rm pr}\{M_j=1\mid \by_j,z_j\}&=& {\rm pr} \{M_j=1\mid \by_j\}\nonumber\\
&=&q(\by_j;\bbeta,\bxi),
\label{eq:g6}
\end{eqnarray}
where the parameter $\bxi$ is distinct from $\bbeta$.
\vs

An obvious choice for the function $q(\by_j;\bbeta,\bxi)$ is the logistic model
\citep{molenberghs_2014_handbook},
\begin{equation}
q(\by_j;\bbeta,\bxi)= \frac{\exp\{\xi_0+\xi_1 e_j\}}
{1+\exp\{\xi_0+\xi_1 e_j\}}
\label{eq:g7}
\end{equation}
where $\bxi=(\xi_0,\xi_1)^T.$
\vs

The expected proportion $\gamma(\bPsi)$ of unclassified features in a 
partially classified sample $\bx_{\rm PC}$ is given by
\begin{eqnarray}
\gamma(\bPsi)&=&\sum_{j=1}^n E(M_j)/n\nonumber\\
&=&E[{\rm pr}\{M_j=1\mid \bY_j\}]\nonumber\\
&=&E\{q(\bY;\bbeta,\bxi)\},
\label{eq:g8}
\end{eqnarray}
where $\bPsi=(\btheta^T,\bxi^T)^T$.

To simplify the numerical computation in the particular case of 
only $g=2$ classes as under consideration, we henceforth replace $e_j$
in (\ref{eq:g7}) by the square of the discriminant function
$d(\by_j;\bbeta)$ to give
\begin{equation}
q(\by_j;\bbeta,\bxi)= \frac{\exp\{\xi_0+\xi_1 d(\by_j;\bbeta)^2\}}
{1+\exp\{\xi_0+\xi_1 d(\by_j;\bbeta)^2\}}.
\label{eq:g9}
\end{equation}
The term $d(\by_j;\bbeta)^2$ can play a similar role 
as the entropy $e_j$ to weight the difficulty in
classifying a feature vector $\by_j$.
More precisely, the square of the value of the discriminant function
$d(\by_j; \bbeta)^2$ is a monotonically decreasing function of the entropy 
$e_j$, and is related to the distance
between a feature vector $\by_j$ and the decision boundary in the feature space. 
Figure 2 shows simulated
data using different parameter values for the missingness mechanism. 
Five hundred values of $\bx$ were simulated from the canonical model 
with $\pi_1=\pi_2$ and $\Delta=2$.
The missingness model was then
applied to the simulated features with $\xi_0 = 3$ and 
$\xi_1 = -0.1, -0.5, -1, -2, -5, -10$. Black squares
denote unclassified features, red triangles are features in Class $C_1$, 
and blue circles are features in $C_2$.
Moving through the Panels (a) to (f), the unclassified features become
more concentrated around the decision boundary as $\xi_1$ decreases. 
The proportion of unclassified features is different in each panel.

The full likelihood function $L_{\rm PC}^{(\rm full)}(\bPsi)$
for $\bPsi$ that can be formed from the partially classified sample 
$\bx_{\rm PC}$ is defined by
\begin{eqnarray}
\log L_{\rm PC}^{(\rm full)}(\bPsi)=
\log L_{\rm PC}^{(\rm ig)}(\btheta)
+\log L_{\rm PC}^{(\rm miss)}(\bbeta,\bxi),
\label{eq:g9a}
\end{eqnarray}
where $\log L_{\rm PC}^{(\rm ig)}(\btheta)$ is defined by (\ref{eq:g4})
and where
$$\log L_{\rm PC}^{(\rm miss)}(\bbeta,\bxi)=
\sum_{j=1}^n [(1-m_j)\log \{1- q(\by_j;\bbeta,\bxi)\}
+ m_j \log q(\by_j;\bbeta,\bxi)]$$
is the log likelihood function for $\bbeta$ and $\bxi$ 
formed on the basis of the missing-label indicators
$m_j\,(j=1,\,\ldots,\,n)$.
\vs
\begin{figure}[H]
    \centering
    \includegraphics[width=\textwidth]{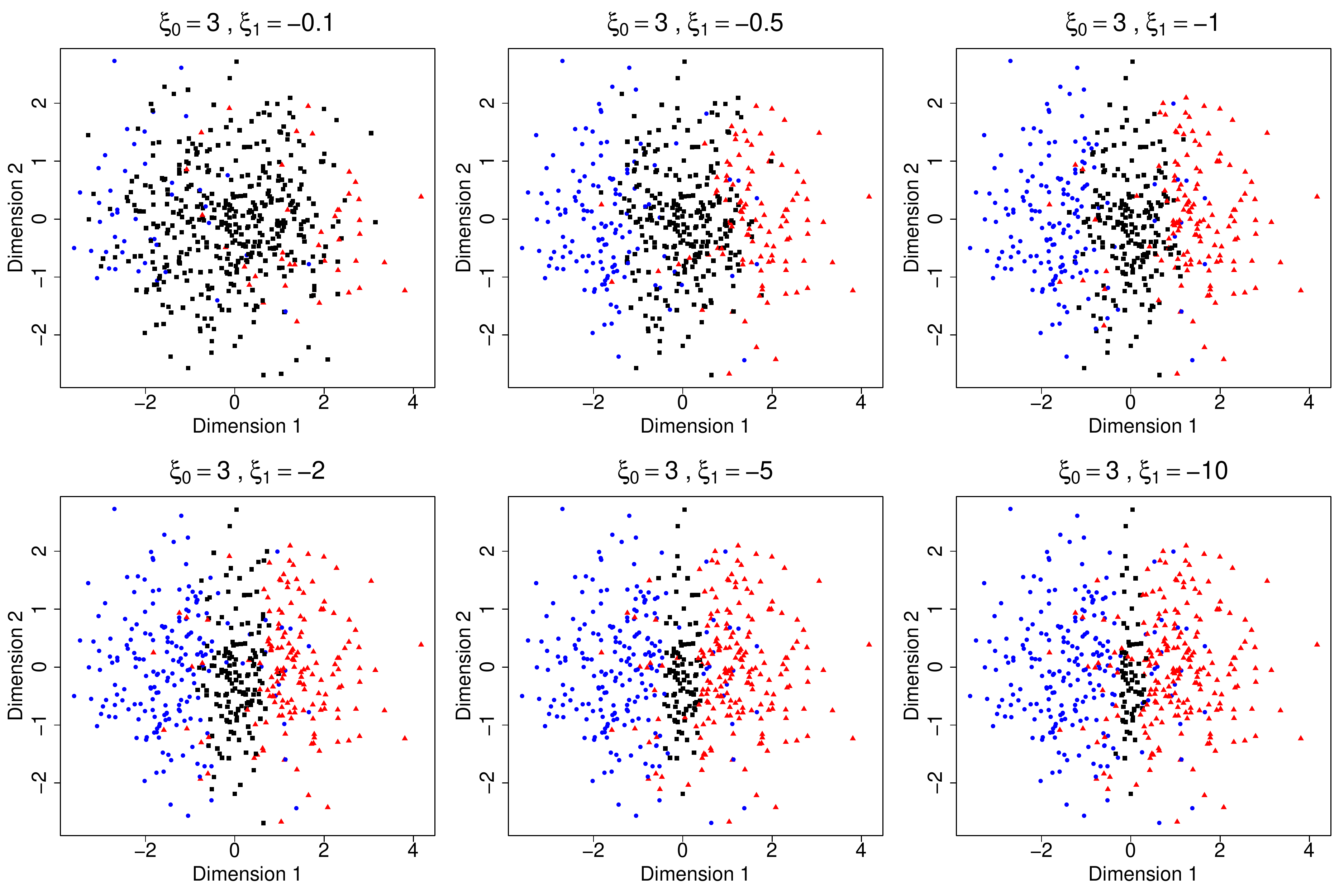}
    \caption{Example simulated datasets using the canonical normal discriminant model and the missingness model (\ref{eq:g9}). 
Here $n=500, \Delta=2, \pi_{1}=\pi_{2}, p=2, \xi_{0}=3$. 
In panels (a) through (f) $\xi_{1}=-0.1, -0.5, 1, -2, -5, -10$, respectively.}
    \label{fig:example_datasets}
\end{figure}

We note that there may be an identifiability issue concerning $\bbeta$
and $\bxi$ if $\log L_{\rm PC}^{(\rm miss)}(\bbeta,\bxi)$ 
given by (\ref{eq:g9})
were to be used on its own for the estimation of $\bbeta$ and $\bxi$. 
But as it is being combined with $\log L_{\rm PC}^{(\rm ig)}(\btheta)$ 
to form the full log likelihood $\log L_{\rm PC}^{(\rm full)}(\bPsi)$,
$\bbeta$ and $\bxi$ are each identifiable
with the use of the latter.
\vs

\section{Fisher Information}
In this section, we derive the Fisher information about $\bbeta$
in the partially classified sample $\bx_{\rm PC}$.
We reparameterize the two-class normal model by taking
\begin{equation}
\btheta=(\btheta_1^T,\bbeta)^T,
\label{eq:g10}
\end{equation}
where $\btheta_1$ contains the elements of $\bmu=\pi_1\bmu_1+\pi_2\bmu_2$
and the distinct elements of
$\bLambda=\bSigma+\pi_1\pi_2(\bmu_1-\bmu_2)
(\bmu_1-\bmu_2)^T$. 
We can now write the vector $\bPsi$ of all unknown parameters,
including the parameter $\bxi$ in the logistic model defined 
by (\ref{eq:g9}), as
\begin{eqnarray}
\bPsi&=&(\btheta^T,\bxi^T)^T\label{eq:g11}\\
&=&(\btheta_1^T,\bbeta^T,\bxi^T)^T.
\label{eq:g12}
\end{eqnarray}

\noindent
{\bf Theorem 1 (Main Result).} {\it The Fisher information about $\bbeta$ 
in the partially classified sample $\bx_{\rm PC}$ 
via the full likelihood function $L_{\rm PC}^{(\rm full)}(\bPsi)$ 
can be decomposed as
\begin{equation}
\bI_{\rm PC}^{(\rm full)}(\bbeta)= \bI_{\rm CC}(\bbeta) 
-\gamma(\bPsi)\bI_{\rm CC}^{(\rm clr)}(\bbeta) + 
\bI_{\rm PC}^{(\rm miss)}(\bbeta),
\label{eq:g12a}
\end{equation}
where
$\bI_{\rm CC}(\bbeta)$ 
is the information about $\bbeta$ 
in the completely classified sample $\bx_{\rm CC},
\bI_{\rm CC}^{(\rm clr)}(\bbeta)$ is the conditional information 
about $\bbeta$ under the logistic regression model 
for the distribution of the class labels given 
the features in $\bx_{\rm CC}$, and
$\bI_{\rm PC}^{(\rm miss)}(\bbeta)$ is the 
information about
$\bbeta$ in the missing-label indicators under the
assumed logistic model for their distribution 
given their associated features 
in the partially classified sample $\bx_{\rm PC}$.
}
\vs

\noindent
{\bf Remark 1.} Since $\gamma(\bPsi)$ is the probability that $M=1$, 
it follows that the second term on the 
right-hand side of (\ref{eq:g12a}),
$\gamma(\bPsi)\,\bI_{\rm CC}^{(\rm clr)}(\bbeta)$,
can be expressed as
$$nE\{(-\partial^2 \log f_{Z\mid Y}(Z\mid\bY;\bbeta)/
\partial\bbeta\partial\bbeta^T)\,q(\bY;\bbeta,\bxi)\},$$
which is the expected information (under the logistic model)
for those class labels $z_j$ in $\bx_{\rm CC}$ 
for which their associated features $\by_j$
would have missing labels $m_j=1$
under the assumed model (\ref{eq:g9})
for missingness. 
\vs

\noindent
{\bf Proof of Theorem 1.}  From the definition (\ref{eq:g9a})
of the full log  likelihood function 
$\log L_{\rm PC}^{(\rm full)}(\bPsi)$,
we can decompose the information matrix 
$\bI_{\rm PC}^{(\rm full)}(\bPsi)$ for $\bPsi$ as
\begin{eqnarray}
\bI_{\rm PC}^{(\rm full)}(\bPsi)
&=&E\{-\partial^2\log L_{\rm PC}^{(\rm full)}(\bPsi)
/\partial\bPsi\partial\bPsi^T\}\nonumber\\
&=&E\{-\partial^2\log L_{\rm PC}^{(\rm ig)}(\btheta)/
\partial\bPsi\partial\bPsi^T\}\nonumber\\
&&\qquad + E\{-\partial^2\log L_{\rm PC}^{(\rm miss)}(\bbeta,\bxi)/
\partial\bPsi\partial\bPsi^T\}\nonumber\\
&=&\bI_{\rm PC}^{(\rm ig)}(\bPsi) + \bI_{\rm PC}^{(\rm miss)}(\bPsi) ,
\label{eq:g13}
\end{eqnarray}
where
\begin{equation}
\bI_{\rm PC}^{(\rm ig)}(\bPsi)=
E\{-\partial^2\log L_{\rm PC}^{(\rm ig)}(\btheta)/\partial\bPsi\partial\bPsi^T\}
\label{eq:g14}
\end{equation}
and
\begin{equation}
\bI_{\rm PC}^{(\rm miss)}(\bPsi)=
E\{-\partial^2\log L_{\rm PC}^{(\rm miss)}(\bbeta,\bxi)/
\partial\bPsi\partial\bPsi^T\}.
\label{eq:g15}
\end{equation}

Considering the first term on the right-hand side 
of (\ref{eq:g13}), we consider its submatrix
\begin{equation}
\bI_{\rm PC}^{(\rm ig)}(\btheta)=
E\{-\partial^2\log L_{\rm PC}^{(\rm ig)}(\btheta)/
\partial\btheta\partial\btheta^T\}.
\label{eq:g16}
\end{equation}
It can be expressed as 
\begin{eqnarray}
&&n [1-\gamma(\bPsi)]
E\{-\partial^2 \log f_{yz}(\bY,Z;\btheta)/\btheta\btheta^T\mid M=0\}\nonumber\\
&&+ n\gamma(\bPsi)
E\{-\partial^2 \log f_y(\bY;\btheta)/\partial\btheta\partial\btheta^T\mid M=1\},
\label{eq:g17}
\end{eqnarray}
where
$f_{yz}(\by,z;\btheta)$ denotes the joint density of $\bY$ and $Z$
and
\begin{equation}
f_y(\by;\btheta)=f_{yz}(\by,z;\btheta)/
f_{z\mid y}(z\mid\by;\bbeta)
\label{eq:g18}
\end{equation}
is the marginal density of $\bY$, and 
where $f_{z\mid y}(z\mid\by;\bbeta)$ is
the conditional probability of $Z$ given $\bY=\by$.

On using (\ref{eq:g18}) in (\ref{eq:g17}), we can write 
$\bI_{\rm PC}^{(\rm ig)}(\btheta)$ as 
\begin{eqnarray*}
\bI_{\rm PC}^{(\rm ig)}(\btheta) &=&
n[1-\gamma(\bPsi)]E\{-\partial^2 
\log f_{yz}(\bY,Z;\btheta)/\partial\btheta\partial\btheta^T\mid M=0\}\\
&&+ n\gamma(\bPsi) E\{-\partial^2 
\log f_{yz}(\bY,Z;\btheta)/\partial\btheta\partial\btheta^T\mid M=1\}\\
&&-n\gamma(\bPsi) E\{-\partial^2 \log f_{z\mid y}(Z\mid\bY;\bbeta)/
\partial\btheta\partial\btheta^T\mid M=1\},
\end{eqnarray*}
which equals
\begin{eqnarray*}
&&nE\{-\partial^2 
\log f_{yz}(\bY,Z;\btheta)/\partial\btheta\partial\btheta^T\}\\
&&-\gamma(\bPsi) n E\{-\partial^2 \log f_{z\mid y}(Z\mid \bY;\btheta)/
\partial\btheta\partial\btheta^T\mid M=1\},
\end{eqnarray*}
and so
\begin{eqnarray}
\bI_{\rm PC}^{(\rm ig)}(\btheta) 
&=&\bI_{\rm CC}(\btheta) -\gamma(\bPsi) \bI_{\rm CC}^{(\rm clr)}(\btheta),
\label{eq:g19}
\end{eqnarray}
where
\begin{eqnarray}
\bI_{\rm CC}(\btheta)&=&
nE\{-\partial^2
\log f_{yz}(\bY,Z;\btheta)/\partial\btheta\partial\btheta^T\}\nonumber\\
&=&E\{-\partial^2\log L_{\rm CC}(\btheta)/\partial\btheta\partial\btheta^T\}
\label{eq:g20}
\end{eqnarray}
is the information about $\btheta$ in the completely classified sample
and where,  corresponding to the partition in (\ref{eq:g10})
of $\btheta$,
\begin{equation}
\bI_{\rm CC}^{(\rm clr)}(\btheta)=
\left(\begin{array}{cc}
\bO&\bO\\
\bO&\bI_{\rm CC}^{(\rm clr)}(\bbeta)
\end{array}\right),
\label{eq:g21}
\end{equation}
since the likelihood function for the logistic regression model does not
contain $\btheta_1$.
Here
\begin{equation}
\bI_{\rm CC}^{(\rm clr)}(\bbeta)=
nE\{-\partial^2 
\log f_{z\mid y}(Z\mid\bY;\btheta)/\partial\bbeta\partial\bbeta^T
\mid M=1\}
\label{eq:g22}
\end{equation}
is the expectation conditional on $M=1$
of the negative Hessian of the conditional density of $Z$ given $\bY$
under the logistic regression model fitted to the completely classified sample.

On considering now the first term on the right-hand side 
of (\ref{eq:g19}), we have that
the information about $\btheta$ in the 
completely classified sample can be partitioned  as
\begin{eqnarray}
\bI_{\rm CC}(\btheta)&=& E\{-\partial^2\log L_{\rm CC}(\btheta)/
\btheta\btheta^T\}\nonumber\\
&=& 
\left(\begin{array}{cc} \bA_{11}&\bA_{12}\\
\bA_{21}&\bA_{22}\end{array}\right),
\label{eq:g23}
\end{eqnarray}
where this partition of $\bI_{\rm CC}(\btheta)$ corresponds to the
partition (\ref{eq:g10}) of $\btheta$.
We partition the inverse of $\bI_{\rm CC}(\btheta)$ as 
$$\bI_{\rm CC}^{-1}(\btheta)=
\left(\begin{array}{cc} \bA^{11}&\bA^{12}\\
\bA^{21}&\bA^{22}\end{array}\right)$$
to give the asymptotic covariance matrix of the 
ML estimator of $\btheta$.

It follows that the information matrix for $\bbeta$
based on the likelihood formed from the completely classified sample
is given by the inverse of $\bA^{22}$,
\begin{eqnarray}
\bI_{\rm CC}(\bbeta)&=&(\bA^{22})^{-1}\nonumber\\
&=& \bA_{22}-\bA_{21}\bA_{11}^{-1}\bA_{12}.
\label{eq:g24}
\end{eqnarray}

As $L_{\rm PC}^{(\rm ig)}(\btheta)$ does not contain $\bxi$, 
it follows from (\ref{eq:g21}) and  (\ref{eq:g24})
that the first term on the right-side of 
(\ref{eq:g19}) for the information matrix
$\bI_{\rm PC}^{(\rm full)}(\bPsi)$ can be partitioned  
corresponding to the partition (\ref{eq:g12}) of $\bPsi$ as
\begin{equation}
\bI_{\rm PC}^{(\rm ig)}(\bPsi)=
\left(\begin{array}{ccc}
\bA_{11}&\bA_{12}&\bO\\
\bA_{21}&\bA_{22}-\gamma(\bPsi)\bI_{\rm CC}^{(\rm clr)}(\bbeta)&\bO\\
\bO&\bO&\bO
\end{array}\right).
\label{eq:g25}
\end{equation}

On considering the other term $\bI_{\rm PC}^{(\rm miss)}(\bPsi)$ 
on the right-hand side of (\ref{eq:g13})
for the information matrix  about $\bPsi$ via the full likelihood
function $L_{\rm PC}^{(\rm full)}(\bPsi)$,
it can be partitioned corresponding to the partition (\ref{eq:g12})
of $\bPsi$ as
\begin{equation}
\bI_{\rm PC}^{(\rm miss)}(\bPsi)=
\left(\begin{array}{ccc} \bO&\bO&\bO\\
\bO&\bB_{22}&\bB_{23}\\
\bO&\bB_{32}&\bB_{33} \end{array}\right),
\label{eq:g26}
\end{equation}
since $L_{\rm PC}^{(\rm miss)}(\bbeta,\bxi)$ does not contain $\btheta_1$.

On using (\ref{eq:g26}) and (\ref{eq:g25})
in (\ref{eq:g13}),
we have that the information matrix $\bI_{\rm PC}^{(\rm full)}(\bPsi)$
for $\bPsi$ on the basis of the full likelihood $L_{\rm PC}^{(\rm full)}(\bPsi)$ 
fitted to the partially classified sample $\bx_{\rm PC}$ 
can be partitioned as 
\begin{eqnarray}
\bI_{\rm PC}^{(\rm full)}(\bPsi)&=&
\left(\begin{array}{ccc} \bA_{11}&\bA_{12}&\bO\\
\bA_{21}&\bA_{22}-\gamma(\bPsi) \bI_{\rm CC}^{(\rm clr)}(\bbeta)+\bB_{22}&\bB_{23}\\
\bO&\bB_{32}&\bB_{33} \end{array}\right).
\label{eq:g28}
\end{eqnarray}
Corresponding to this partition of $\bI_{\rm PC}^{(\rm full)}(\bPsi)$,
we write it as
\begin{equation}
\bH=
\left(\begin{array}{ccc}
\bH_{11}&\bH_{12}&\bH_{13}\\
\bH_{21}&\bH_{22}&\bH_{23}\\
\bH_{31}&\bH_{32}&\bH_{33}\\
\end{array}\right),
\label{eq:gg25}
\end{equation}
and we let $\bH^{ij}$ denote the block in $\bH^{-1}$ corresponding
to the block $\bH_{ij}$ in $\bH\,(i,j=1,2,3)$. 

The inverse of the matrix $\bH^{22}$ provides the information matrix
$\bI_{\rm PC}^{(\rm full)}(\bbeta)$ for $\bbeta$,
where $\bbeta$ is estimated by consideration
of the full likelihood function $L_{\rm PC}^{(\rm full)}(\bPsi)$.
To calculate $\bH^{22}$, we refine the partition (\ref{eq:gg25})
of $\bH$ to 
\begin{equation}
\bH=
\left(\begin{array}{cc}
\bW_{11}&\bW_{12}\\
\bW_{21}&\bW_{22}\\
\end{array}\right),
\label{eq:gg26}
\end{equation}
where
\begin{eqnarray*}
\bW_{11}&=&
\left(\begin{array}{cc}
\bH_{11}&\bH_{12}\\
\bH_{21}&\bH_{22}\\
\end{array}\right),\quad \bW_{12}=\bW_{21}^T=\begin{pmatrix}
        \bzero\\
        \bB_{23}\end{pmatrix},
\end{eqnarray*}
and $\bW_{22}=\bB_{33}$.
Using standard results for the inversion of matrices in block form,
we have that
\begin{eqnarray}
\left(\begin{array}{cc}
\bH^{11}&\bH^{12}\\
\bH^{21}&\bH^{22}
\end{array}\right)&=&
(\bW_{11}-\bW_{12}\bW_{22}^{-1}\bW_{21})^{-1}\nonumber\\
&=&
\left(\begin{array}{cc}
\bA_{11}&\bA_{12}\\
\bA_{21}&\bH_{22}-\bB_{23}\bB_{33}^{-1}\bB_{32}
\end{array}\right)^{-1}.
\nonumber\\
\label{eq:gg27}
\end{eqnarray}

Now $\bI_{\rm PC}^{(\rm full)}(\bbeta)=\{\bH^{22}\}^{-1}$, which 
can be calculated from (\ref{eq:gg27})
to give
\begin{eqnarray*}
\{\bH^{22}\}^{-1}&=&
\bH_{22} -\bB_{23}\bB_{33}^{-1}\bB_{32}
-\bA_{21}\bA_{11}^{-1}\bA_{12}\\
&=&(\bA_{22}-\bA_{21}\bA_{11}^{-1}\bA_{12})
-\gamma(\bPsi)\bI_{\rm PC}^{\rm clr)}(\bbeta)\\
&=&\quad +\, (\bB_{33}-\bB_{23}\bB_{33}^{-1}\bB_{32})\\
&=&\bI_{\rm CC}(\bbeta) -\gamma(\bPsi)\bI_{\rm Psi}^{(\rm clr)}(\bbeta)
+\bI_{\rm PC}^{(\rm miss)}(\bbeta),
\end{eqnarray*}
on noting (\ref{eq:g24}) and that
$$\bI_{\rm PC}^{(\rm miss)}(\bbeta) = \bB_{22}-\bB_{23}\bB_{33}^{-1}\bB_{32}$$
is the information about $\bbeta$  in the missing-label indicators.
\vs

\noindent
{\bf Remark 2.} Note that the contribution $\bI_{\rm PC}^{(\rm miss)}(\bbeta)$
to the full information matrix
would be equal to $\bB_{22}$  if $\bxi$ were known, so the term
$$\bB_{23}\bB_{33}^{-1}\bB_{32}$$
can be viewed as the loss of information about $\bbeta$ by virtue
of $\bxi$ not being known 
and having to be estimated as well as $\bbeta$.
\vs

\section{Asymptotic Relative Efficiencies}
We let 
(i) $\hbbeta_{\rm CC}$ denote the maximum likelihood (ML) estimate of
$\bbeta$ by consideration of the likelihood function $L_{\rm CC}(\btheta)$
that can be formed from the
completely classified sample $\bx_{\rm CC}$;
(ii) $\hbbeta_{\rm PC}^{(\rm ig)}$ denote
the ML estimate of $\bbeta$ on the basis of the likelihood function
$L_{\rm PC}^{(\rm ig)}(\btheta)$
formed from the partially classified sample
$\bx_{\rm PC}$ by ignoring the missingness in the labels
of the unclassified features;
(iii) $\hbbeta_{\rm PC}^{(\rm full)}$ denote the ML estimate of $\bbeta$ 
by consideration of the full likelihood function 
$L_{\rm PC}^{(\rm full)}(\bPsi)$.
\vs

We let
$\hR_{\rm CC}, \hR_{\rm PC}^{(\rm ig)},$ and
$\hR_{\rm PC}^{(\rm full)}$
denote the estimated Bayes' rule obtained by plugging in
the estimates
$\hbbeta_{\rm CC}, \hbbeta_{\rm PC}^{(\rm ig)}$, and
$\hbbeta_{\rm PC}^{(\rm full)}$, respectively, for $\bbeta$ in 
the Bayes' rule $R(\by;\bbeta).$
\vs

The overall error rate of the Bayes' rule $R(\by;\bbeta)$ is denoted by 
${\rm err}(\bbeta)$ (the optimal error rate). 
The conditional error rates of the estimated Bayes' rules
$\hR_{\rm CC},
\hR_{\rm PC}^{\rm (ig)}$,
and $\hR_{\rm PC}^{(\rm full)}$ are denoted by
${\rm err}(\hbbeta_{\rm CC}),
{\rm err}(\hbbeta_{\rm PC}^{(\rm ig)}),$
and ${\rm err}(\hbbeta_{\rm PC}^{(\rm full)}),$
respectively.
The asymptotic relative efficiency (ARE) of the rule
$\hR_{\rm PC}^{(\rm full)}$ compared to the rule 
$\hR_{\rm CC}$ 
based on the completely classified sample 
is defined as
\begin{equation}
{\rm ARE}(\hR_{\rm PC}^{(\rm full)})
=\frac{E\{{\rm err}(\hbbeta_{\rm C})\} -{\rm err}(\bbeta)}
{E\{{\rm err}(\hbbeta_{\rm PC}^{(\rm full)}\}-{\rm err}(\bbeta)},
\label{eq:g34}
\end{equation}
where the expectation in the numerator and denominator of (\ref{eq:g34}) 
is taken over the distribution of the estimators of $\bbeta$
and is expanded up to terms of the first order.
\vs

Under the assumption that the class labels are missing completely 
at random, \citet{ganesalingam_1978_efficiency} derived the 
ARE of $\hR_{\rm PC}^{(\rm ig)}$
compared to $\hR_{\rm CC}$,
$${\rm ARE}(\hR_{\rm PC}^{(\rm ig)})
=\frac{E\{{\rm err}(\hbbeta_{\rm CC})\} -{\rm err}(\bbeta)}
{E\{{\rm err}(\hbbeta_{\rm PC}^{(\rm ig)}\}-{\rm err}(\bbeta)},$$
in the case of a completely unclassified sample $(\gamma = 1)$ 
for univariate features $(p = 1)$.
Their results are listed in Table 1 
for $\Delta = 1, 2,$ and 3. 
\citet{oneill_1978_normal} extended their result to
multivariate features and for arbitrary $\gamma$.
His results showed that this ARE was not sensitive to the
values of $p$ and does not vary with $p$ for equal class
prior probabilities. 
Not surprisingly, it can be seen from Table \ref{tab:mcar}
that the ARE of $\hR_{\rm PC}^{(\rm ig)}$
for a totally unclassified
sample is low, particularly for classes weakly separated as 
represented by $\Delta= 1$ in Table~1.
\vs

\begin{table}[]
\centering
\begin{tabular}{@{}lllll@{}}
\toprule
$\pi_{1}$ & $\Delta=1$ & $\Delta=2$ & $\Delta=3$ & $\Delta=4$ \\ \midrule
0.1       & 0.0036       & 0.0591       & 0.2540      & 0.5585      \\
0.2       & 0.0025       & 0.0668       & 0.2972      & 0.6068      \\
0.3       & 0.0027       & 0.0800       & 0.3289      &0.6352      \\
0.4       & 0.0038       & 0.0941       & 0.3509      & 0.6522      \\
0.5       & 0.0051       & 0.1008      & 0.3592      & 0.6580      \\ \bottomrule
\end{tabular}
\caption{Asymptotic relative efficiency of $\hR_{\rm PC}^{(\rm ig)}$ 
compared to $\hR_{\rm CC}$}
\label{tab:mcar}
\end{table}

In other work on the ARE of $\hR_{\rm PC}^{(\rm ig)}$ compared to $\hR_{\rm CC}$,
\citet{mclachlan1995asymptotic}  evaluated it where the unclassified
univariate features had labels missing at random (MAR) due to 
truncation of the features.
\vs

Here the focus is on the ARE of the $\hR_{\rm PC}^{(\rm full)}$
where additional information on $\bbeta$
from the missing-data mechanism is incorporated 
into the full likelihood function to yield the
full ML estimator $\hbbeta_{\rm PC}^{(\rm full)}$
on the basis of the partially classified sample $\bx_{\rm PC}$.
\vs

We now sketch the derivation of the ARE of $\hR_{\rm PC}^{(\rm full)}$.
We let $\hbbeta$ denote a generic estimator of $\bbeta$ that satisfies
\begin{equation}
\sqrt{n} 
\left(\begin{array}{c}
\hbeta_0-\beta_0\\
\hbbeta_1-\bbeta_1
\end{array}\right) \stackrel{\cD} \rightarrow  N(\vect{0},\bV),
\label{eq:g35}
\end{equation}
as $n \rightarrow \infty$,
and that the first and second moments also converge.
Then the first order expansion of the so-called
excess error rate, that is,
the expected error rate ${\rm err}(\hbbeta)$
over the optimal rate ${\rm err}(\bbeta)$ for the estimated Bayes' rule
$R(\by;\hbbeta)$, can be expanded as 
\begin{equation}
E\{{\rm err}(\hbbeta)\}- {\rm err}(\bbeta)=
n^{-1}\,{\rm tr}(\bJ\bV) + o(1/n),
\label{eq:g36}
\end{equation}
where
$$\bJ={\textstyle\frac{1}{2}}
[\nabla\nabla^T {\rm err}(\hbbeta)]_{\hbbeta=\bbeta}$$
and $\nabla=
(\partial/\partial\hbeta_1,\,\ldots,\,\partial/\partial\hbeta_p)^T.$
\vs

In deriving the ARE of logistic regression,
\citet{efron_1975_efficiency} showed 
under the canonical for (\ref{eq:g2b}) adopted here 
for the two-class normal discrimination model 
that the expansion (\ref{eq:g36}) reduces to
\begin{equation}
E\{{\rm err}(\hbbeta)\}-{\rm err}(\bbeta)=\frac{\pi_1\phi(\Delta^*;0,1)}{2\Delta n}
[v_{00}-\frac{2}{\Delta}\lambda v_{01}
+\frac{\lambda^2}{\Delta^2}v_{11}
+ v_{22}+ ...+ v_{pp}]+o(1/n),
\label{eq:g37}
\end{equation}
where $\lambda=\log(\pi_1/\pi_2),
\Delta^*={\textstyle\frac{1}{2}}\Delta-\lambda/\Delta$, and
$\phi(y;\mu,\sigma^2)$ denotes
the normal density with mean $\mu$ and variance $\sigma^2$.
Here $v_{jk}=(\bV)_{jk}$, where
the columns and rows in $\bV$ are indexed from zero to $p$.
\vs

The following theorem gives the ARE of 
$\hR_{\rm PC}^{(\rm full)}$ compared to $\hR_{\rm CC}$
in the case of equal prior probabilities $\pi_1=\pi_2$.
\vs

\noindent
{\bf Theorem 2.} {\it Under the missing-label model defined by
(\ref{eq:g9}), 
the ARE of $\hR_{\rm PC}^{(\rm full)}$ 
compared to $\hR_{\rm CC}$ is given in the case of $\pi_{1}=\pi_{2}$ by
\begin{equation}
{\rm ARE}(\hR_{\rm PC}^{(\rm full)})=4(1+\Delta^2/4)u_0
\label{eq:g38}
\end{equation}
for all $p$, where 
\begin{align}
   u_0 &= 1/\{4(1+\Delta^2/4)\}-\gamma d_0+b_0, \label{eq:g39} \\
   b_0&= \int_{-\infty}^{\infty}4\xi_1^2\Delta^2y_1^2q_1(y_1)
(1-q(y_1))f_{y_{1}}(y_1)dy_1,\nonumber \\
d_0&= \int_{-\infty}^{\infty}\tau_1(y_1)\tau_2(y_1)q_1(y_1)
\gamma^{-1}f_{y_1}(y_1)dy_1, \nonumber
\end{align}
and where 
\begin{align*}
\tau_1(y_1)&={\rm pr}\{Z=1\mid (\bY)_1=y_1\}\quad (i=1,2),\\
q_1(y_1;\Delta,\bxi)&={\rm pr}\{M=1\mid (\bY)_1=y_1\},\\
f_{y_1}(y_1;\Delta,\pi_1)&=\pi_1\phi(y_1;\Delta/2,1)
+(1-\pi_1)\phi(y_1;-\Delta/2,1).
\end{align*}
}
\vs

In the above definitions of $b_0$ and $d_0$, we have 
suppressed the dependence of $\tau_1(y_1), q_1(y_1)$, and $f_{y_1}(y_1)$ 
on $\Delta, \pi_1,$ and $\bxi$.
\vs

\noindent
{\bf Proof of Theorem 2.}
To derive the ARE of $\hR_{\rm PC}^{(\rm full)}$, we have to calculate
the first order expansions of the numerator and denominator of 
the right-hand side of (\ref{eq:g34}).
Now the first order expansion of the numerator of (\ref{eq:g34}) has 
been given by \citet{efron_1975_efficiency} for arbitrary values of
$\pi_1, \Delta$, and $p$ under the adopted canonical form. 
It is given for $\pi_{1}=\pi_{2}$ by
\begin{align}
    E\{{\rm err}(\hbbeta_{\rm CC})\}
-{\rm err}(\bbeta)&=\frac{p\phi(\Delta/2;0,1)
(1 + \Delta^2/4)}{\Delta n}+  o(1/n). 
\label{eq:err_numerator}
\end{align}

To obtain the denominator of (\ref{eq:g34}) under the
adopted canonical form, we apply the following result 
of \cite[Theorem 1]{efron_1975_efficiency}, 
who developed it in the course of deriving the 
ARE of logistic regression 
under the canonical form (\ref{eq:g2b})
adopted here for the two-class normal
discrimination model.

Let $\hbbeta$ be an estimator of $\bbeta$ for which
$\sqrt{n}(\hbbeta-\bbeta)$ converges in distribution 
to the $N(\bzero,\bV)$ distribution, 
as $n \rightarrow \infty$, 
and that the first and second order moments also converge.
Then the expectation of the so-called excess error rate 
can be expanded as 
\begin{eqnarray}
E\{{\rm err}(\hbbeta)\}-{\rm err}(\bbeta)
&=&\frac{\pi_1\phi(\Delta^*;0,1)}{2\Delta n}\,w + o(1/n),
\label{eq:g37}
\end{eqnarray}
where
$$w=v_{00}-\frac{2\lambda}{\Delta} v_{01}
+\frac{\lambda^2}{\Delta^2}v_{11}
+\sum_{i=2}^p  v_{ii} $$
and where $\lambda=\log(\pi_1/\pi_2), 
\Delta^*={\textstyle\frac{1}{2}}\Delta-\lambda/\Delta$, and 
$\phi(y;\mu,\sigma^2)$ denotes 
the normal density with mean $\mu$ and variance $\sigma^2$.
Here $v_{jk}=(\bV)_{jk}$, where
the columns and rows in $\bV$ are indexed from zero to $p$.

In order to apply the result (\ref{eq:g37}) for
$\hbbeta$ equal to the full ML estimator 
$\hbbeta_{\rm PC}^{(\rm full)}$ of $\bbeta,$ 
we need to invert $(1/n)$ times the information matrix 
$\bI_{\rm PC}^{(\rm full)}(\bbeta)$
for $\bbeta$ given by (\ref{eq:g12a}) in Theorem 1.
This evaluation is simplified in the case of $\pi_1=\pi_2$ on noting 
several of the submatrices of the matrices
in (\ref{eq:g12a}) become diagonal.
On inverting $\bI_{\rm PC}^{(\rm full)}(\bbeta)$, we find that when $\pi_{1}=\pi_{2}$,
\begin{eqnarray}
v_{jj}&=&1/u_0, \quad  j \neq 1, \label{eq:g41}
\end{eqnarray}
where $u_0$ is defined by (\ref{eq:g39}).
Substituting into (\ref{eq:g41}), it follows that for $\pi_{1}=\pi_{2}$,
\begin{eqnarray} 
E\{{\rm err}(\hbbeta_{\rm PC}^{(\rm full)})\}
-{\rm err}(\bbeta)
&=& \frac{p\phi(\Delta/2;0,1)}{4n\Delta u_0} + o(1/n),
\label{eq:err_denominator}
\end{eqnarray}
where we have used the fact that $\lambda=0$ when $\pi_{1}=\pi_{2}$.
The ratio of the right-hand side of (\ref{eq:err_numerator}) 
to that of \eqref{eq:err_denominator} 
gives the ARE. 
This completes the proof of Theorem 2.
The extension of this theorem to the case of unequal prior
probabilities is given in the Appendix.

In the case of $\pi_1=\pi_2$,
Table \ref{tab:theoretical} gives the ARE
of $\hR_{\rm PC}^{(\rm full)}$ compared to $\hR_{\rm CC}$
for various combinations of the parameters 
$\Delta, \xi_0$, and $\xi_1$,
the results applying for all values of $p$.
It can be seen for most of the combinations 
in Table \ref{tab:theoretical} that 
the ARE of $\hR_{\rm PC}^{(\rm full)}$
is greater than one,
being appreciably greater than one 
for some combinations of the parameters.
For example, for $\Delta=1$ (representing classes close together)
or $\Delta=2$ (classes moderately separated), the ARE is not less than
15.48 for any combination with $\xi_0$ =2 or 3 and $\xi_1$=
-5 or -10, being as high as 40.4 for
$\Delta=1, \xi_0=5, \xi_1=-10$.
This shows
that the asymptotic expected excess error rate 
using the partially classified sample $\bx_{\rm PC}$
can be much lower than the corresponding excess rate 
using the completely classified sample $\bx_{\rm CC}$.
The contribution to the Fisher information 
from the missingness mechanism can
be relatively very high if $|\xi_{1}|$ is large, 
as the location of the unclassified features 
in the feature space provides information 
about regions of high uncertainty, 
and hence where the absolute value of the discriminant function
$| d(\by_j; \hbbeta)|$ 
should be small. 
Consistent with this, it can be seen in Table 1 that
as $\xi_{1}$ decreases, the ARE of $\hR_{\rm PC}^{(\rm full)}$
increases  
for fixed $\xi_{0}$ and $\Delta$.

In the  Appendix, we give the general expression for the 
ARE of $\hR_{\rm PC}^{(\rm full)}$ for $\pi_1\neq \pi_2.$
We find
that this ARE is not sensitive to the value $\pi_1$ 
in the range (0.2, 0.8), so that Theorem 2 can
provide useful guidelines for arbitrary prior probabilities.

\begin{table}
\centering
\resizebox{\linewidth}{!}{%
\begin{tabular}{@{}llllllllllllllll@{}}
\toprule
         & \multicolumn{5}{c}{$\xi_{0}=1.5$} & \multicolumn{5}{c}{$\xi_{0}=3$} & \multicolumn{5}{c}{$\xi_{0}=5$} \\ \cmidrule(l){2-6} \cmidrule(l){7-11}\cmidrule(l){12-16}
$\Delta$ & 1   & 1.5   & 2   & 2.5   & 3   &    1   & 1.5  & 2   & 2.5  & 3  & 1   & 1.5   & 2   & 2.5  & 3    \\ \midrule
$\xi_{1}=-0.1$ & 0.2 & 0.4 & 0.8 & 1.3 & 1.6 & 0.1 & 0.2 & 0.5 & 1.2 & 1.9 & 0.01 & 0.1 & 0.3 & 0.9 & 1.9 \\ 
$\xi_{1}=-0.5$  & 1.5 & 2.6 & 3.1 & 3.2 & 2.9 & 1.0 & 2.7 & 4.0 & 4.3 & 4.1 & 0.4 & 2.2 & 4.4 & 5.5 & 5.5 \\ 
$\xi_{1}=-1$  & 3.6 & 4.7 & 4.7 & 4.2 & 3.6 & 3.5 & 5.8 & 6.4 & 5.9 & 5.1 & 2.4 & 6.1 & 7.8 & 7.8 & 6.9 \\ 
$\xi_{1}=-5$ & 15.0 & 12.5 & 10.3 & 8.4 & 6.6 & 20.2 & 17.7 & 14.8 & 12.1 & 9.4 & 23.4 & 22.5 & 19.4 & 16.0 & 12.5 \\ 
$\xi_{1}=-10$  & 23.1 & 17.9 & 14.4 & 11.5 & 8.9 & 32.5 & 25.8 & 20.9 & 16.6 & 12.8 & 40.4 & 33.6 & 27.5 & 22.0 & 16.9 \\ 
 \bottomrule
\end{tabular}
}
\caption{ Asymptotic relative efficiency of $\hR_{\rm PC}^{(\rm full)}$
for $\pi_1=\pi_2$ (applicable for all $p$)}
\label{tab:theoretical}
\end{table}

\vs

\begin{table}[H]
\centering
\resizebox{\linewidth}{!}{%
\begin{tabular}{@{}llllllllllllllll@{}}
\toprule
         & \multicolumn{5}{c}{$\xi_{0}=1.5$} & \multicolumn{5}{c}{$\xi_{0}=3$} & \multicolumn{5}{c}{$\xi_{0}=5$} \\ \cmidrule(l){2-6} \cmidrule(l){7-11}\cmidrule(l){12-16}
$\Delta$ & 1   & 1.5   & 2   & 2.5   & 3   &    1   & 1.5  & 2   & 2.5  & 3  & 1   & 1.5   & 2   & 2.5  & 3    \\ \midrule
$\xi_{1}=-0.1$  & 0.80 & 0.75 & 0.66 & 0.53 & 0.38 & 0.95 & 0.93 & 0.87 & 0.75 & 0.58 & 0.99 & 0.99 & 0.97 & 0.91 & 0.77 \\ 
$\xi_{1}=-0.5$  & 0.70 & 0.53 & 0.37 & 0.24 & 0.15 & 0.89 & 0.74 & 0.54 & 0.36 & 0.23 & 0.98 & 0.88 & 0.69 & 0.49 & 0.32 \\ 
 $\xi_{1}=-1$ & 0.60 & 0.41 & 0.27 & 0.17 & 0.10 & 0.80 & 0.58 & 0.39 & 0.25 & 0.15 & 0.93 & 0.74 & 0.52 & 0.34 & 0.21 \\ 
 $\xi_{1}=-5$  & 0.33 & 0.20 & 0.12 & 0.07 & 0.04 & 0.47 & 0.29 & 0.18 & 0.11 & 0.06 & 0.61 & 0.38 & 0.24 & 0.15 & 0.09 \\ 
$\xi_{1}=-10$  & 0.24 & 0.14 & 0.08 & 0.05 & 0.03 & 0.35 & 0.21 & 0.13 & 0.08 & 0.05 & 0.46 & 0.27 & 0.17 & 0.10 & 0.06 \\ 
 \bottomrule
\end{tabular}
}
\caption{Probability of a missing label $\gamma(\bPsi)$ for $\pi_1=\pi_2$}
\label{tab:gamma}
\end{table}

In Table \ref{tab:gamma}, we have listed the probability of a missing label for each combination of the parameters 
in Table~\ref{tab:theoretical}. 
If a feature $\by_j$ is on the decision boundary, then $d(\by_j;\bbeta) =0$
and the conditional probability of a missing label is equal to
$${\rm pr}\{M_j=1 \mid \by_j\}=1/\{1+\exp(-\xi_0)\}.$$
This probability is equal to 0.82, 0.95, and 0.99 for $\xi_0$ = 1.5, 3, and 5, respectively, 
which are the values of $\xi_0$ used in Table 2.
\vs

\section{Simulations}
We conducted a simulation to assess to what extent the asymptotic results
of the previous section apply in practice.
For each of the combinations of the parameters in Table 1, 
we generated $B = 1000$ samples of 
$\bX=(\bY^T \hspace{-0.25em},Z)^T$ to form the
completely classified sample $\bx_{\rm CC}$ and the
partially classified sample $\bx_{\rm PC}$.
On each replication,
the estimates 
$\hbbeta_{\rm CC}$ and $\hbbeta_{\rm PC}^{(\rm full)}$ 
were computed using a quasi-Newton algorithm, along with
the conditional error rates,
${\rm err}(\hbbeta_{\rm CC})$ and ${\rm err}(\hbbeta_{\rm PC}^{(\rm full)}).$
We let ${\rm err}(\hbbeta_{\rm CC}^{(b)})$ and 
${\rm err}(\hbbeta_{\rm PC}^{({\rm full}, b)})$ 
denote the conditional error rate  of $\hR_{\rm CC}$ and of
$\hR_{\rm PC}^{(\rm full)}$, respectively, on the $b$th replication.
The relative efficiency (RE) of $\hR_{\rm PC}^{(\rm full)}$
compared to $\hR_{\rm CC}$ was estimated by
\begin{equation}
\overline{{\rm RE}}(\hR_{\rm PC}^{(\rm full)})=
\frac{
B^{-1}{\sum_{b=1}^B \{{\rm err}(\hbbeta_{\rm CC}^{(b)})-{\rm err}(\bbeta)}\}}
{B^{-1}{\sum_{b=1}^B 
\{{\rm err}(\hbbeta_{\rm PC}^{({\rm full}, b)})-{\rm err}(\bbeta)}\}}.
\label{eq:g42}
\end{equation}
The nonparametric
bootstrap with 1000 resamples was used to assess 
the variability of the estimates \citep{efron_1986_bootstrap}.

Tables \ref{tab:n=100} and \ref{tab:n=500} report the results with the bootstrap 
standard errors in parentheses. 
It can be seen in the case of $n=500$ that there is 
very close agreement between the ARE of
$\hR_{\rm PC}^{(\rm full)}$ and its simulated values for the various combinations
of $\Delta, \xi_0$, and $\xi_1$  in Table \ref{tab:n=500}.
As one would expect, the agreement is not as close 
for the smaller sample size $n=100$,
but there is still good agreement for most of the combinations of the parameters
in Table \ref{tab:n=100}.
The simulated value of the ARE of $\hR_{\rm PC}^{(\rm full)}$ for $n=100$
is less than its actual value for nearly all of the combinations in
Table \ref{tab:n=100} with $\xi_1\leq -0.5.$, indicating that the gain in efficiency
for finite samples is not as high 
as given asymptotically for these combinations. 
One of them for which the agreement between 
the ARE of $\hR_{\rm PC}^{(\rm full)}$  and its simulated value 
is not close is $\Delta =3$ with $\xi_0=3, \xi_1=-10$, 
where the ARE is 12.8 but its simulated value is 4.4. 
A possible explanation for this is that for this combination of the
parameters the probability $\gamma$ that a feature vector will have a missing label 
is very low at 0.06, so in a sample of size $n=100$ the estimation
of $\xi_1$  has to be based on a sample with few values of the 
missing-label indicator variable equal to 1.

\section{Discussion}
The analysis of partially classified data often involves
additional considerations relative to 
completely classified data; see, for example, \citet{chapelle_2010_semi}.
Partially classified data can arise in situations where classifications are made 
by subjective judgement, and there is uncertainty 
on the best assignment for a number of instances in the training set. 
From a statistical point of view, the propensity of high entropy 
features to remain unclassified represents an 
extra source of information for learning a classification rule. 
More formally, the Fisher information in a partially classified 
sample will include a contribution from the missing data mechanism 
under mild assumptions \citep{rubin_1976}. 
We have shown that in the case of two-class 
normal discriminant analysis, the Fisher information
about the vector of discriminant function coefficients
in the partially classified dataset can be much
greater than in a completely classified dataset 
where  the relationship between classification difficulty
and the probability of a missing label is strong. 
As a consequence, 
the asymptotic expected error rate of the classifier
trained using $\bx_{\rm PC}$ can be smaller then the expected error rate 
of the classifier trained using $\bx_{\rm CC}$.
We observed this theoretical superefficiency
in our Monte Carlo simulations. 
We have focused on a simple logistic selection model 
to give mathematical insight into this phenomenon. 
Generic model checking and diagnostic tools can be used 
to assess the goodness of fit of a proposed missingness model. 
Further work will involve the mathematical and empirical study
of more complex models. The distance of unlabelled observations from 
the separating hyperplane has also been identified 
as an important quantity for semi-supervised learning 
with support vector machines \citep{vapnik_1998_statistical}, 
and this is a possible direction to follow to extend 
the proposed methodology to nonlinear models. 
The likelihood contribution of the missingness model can also be viewed 
as a regularisation term that includes the unlabelled observations, 
placing it within a general paradigm in 
semi-supervised learning \citep{berthelot_2019_mixmatch}. 
This perspective may also help to understand the behaviour 
of the full likelihood if the missingness mechanism is misspecified.

\begin{landscape}
\begin{table}
\centering
\resizebox{\linewidth}{!}{%
\begin{tabular}{@{}llllllllllllllll@{}}
\toprule
         & \multicolumn{5}{c}{$\xi_{0}=1.5$} & \multicolumn{5}{c}{$\xi_{0}=3$} & \multicolumn{5}{c}{$\xi_{0}=5$} \\ \cmidrule(l){2-6} \cmidrule(l){7-11}\cmidrule(l){12-16}
$\Delta$ & 1   & 1.5   & 2   & 2.5   & 3   &    1   & 1.5  & 2   & 2.5  & 3  & 1   & 1.5   & 2   & 2.5  & 3    \\ \midrule
$\xi_{1}=-0.1$  & 0.2 (0.01) & 0.3 (0.02) & 0.7 (0.04) & 1 (0.06) & 1.6 (0.09) & 0.1 (0.01) & 0.1 (0.005) & 0.3 (0.04) & 1.2 (0.07) & 1.5 (0.1) & 0.05 (0.01) & 0.01 (0.01) & 0.06 (0.01) & 0.8 (0.05) & 2.0 (0.1) \\
\rowcolor{Gray}$\xi_{1}=-0.1$ & 0.2 & 0.4 & 0.8 & 1.3 & 1.6 & 0.1 & 0.2 & 0.5 & 1.2 & 1.9 & 0.01 & 0.1 & 0.3 & 0.9 & 1.9 \\
$\xi_{1}=-0.5$  & 0.9 (0.1) & 2.1 (0.1) & 2.9 (0.2) & 2.9 (0.2) & 2.6 (0.2) & 0.2 (0.02) & 2.5 (0.2) & 3.7 (0.2) & 4.3 (0.3) & 3.1 (0.2) & 0.1 (0.01) & 0.9 (0.2) & 3.8 (0.2) & 4.9 (0.3) & 4.8 (0.3) \\
\rowcolor{Gray}$\xi_{1}=-0.5$  & 1.5 & 2.6 & 3.1 & 3.2 & 2.9 & 1.0 & 2.7 & 4.0 & 4.3 & 4.1 & 0.4 & 2.2 & 4.4 & 5.5 & 5.5 \\
$\xi_{1}=-1$  & 3.5 (0.2) & 4.6 (0.3) & 4.4 (0.3) & 3.7 (0.2) & 3.0 (0.2) & 1.8 (0.3) & 5.5 (0.4) & 6.5 (0.4) & 5.8 (0.4) & 4.0 (0.3) & 0.2 (0.02) & 5.3 (0.3) & 6.8 (0.4) & 6.6 (0.5) & 5.4 (0.4) \\
\rowcolor{Gray}$\xi_{1}=-1$  & 3.6 & 4.7 & 4.7 & 4.2 & 3.6 & 3.5 & 5.8 & 6.4 & 5.9 & 5.1 & 2.4 & 6.1 & 7.8 & 7.8 & 6.9 \\
$\xi_{1}=-5$  & 14.8 (1) & 10.7 (0.6) & 8.1 (0.5) & 5.9 (0.4) & 3.5 (0.2) & 19.3 (1) & 16.8 (1) & 10.6 (0.7) & 7.4 (0.5) & 4.5 (0.3) & 23.6 (2) & 17 (1) & 13.2 (0.9) & 8.4 (0.6) & 4.6 (0.3) \\
\rowcolor{Gray}$\xi_{1}=-5$ & 15.0 & 12.5 & 10.3 & 8.4 & 6.6 & 20.2 & 17.7 & 14.8 & 12.1 & 9.4 & 23.4 & 22.5 & 19.4 & 16.0 & 12.5 \\
$\xi_{1}=-10$  & 22.1 (2) & 16.5 (1) & 11.5 (0.7) & 6.7 (0.5) & 3.2 (0.2) & 33.3 (2) & 19.9 (1) & 15.2 (1) & 8.8 (0.6) & 4.2 (0.3) & 39.6 (3) & 25.4 (2) & 16.5 (1) & 8.1 (0.5) & 4.2 (0.3) \\
\rowcolor{Gray}$\xi_{1}=-10$  & 23.1 & 17.9 & 14.4 & 11.5 & 8.9 & 32.5 & 25.8 & 20.9 & 16.6 & 12.8 & 40.4 & 33.6 & 27.5 & 22.0 & 16.9 \\
 \bottomrule
\end{tabular}
}
\caption{Simulated relative efficiency of $\hR_{\rm PC}^{(\rm full)}$
with $\pi_1=\pi_2$ for $n=100, p=1$ (Bootstrap standard errors
are in parentheses). The grey shaded rows give for ease of comparison
the values of the ARE from Table 2}
\label{tab:n=100}
\end{table}

\begin{table}
\centering
\resizebox{\linewidth}{!}{%
\begin{tabular}{@{}llllllllllllllll@{}}
\toprule
         & \multicolumn{5}{c}{$\xi_{0}=1.5$} & \multicolumn{5}{c}{$\xi_{0}=3$} & \multicolumn{5}{c}{$\xi_{0}=5$} \\ \cmidrule(l){2-6} \cmidrule(l){7-11}\cmidrule(l){12-16}
$\Delta$ & 1   & 1.5   & 2   & 2.5   & 3   &    1   & 1.5  & 2   & 2.5  & 3  & 1   & 1.5   & 2   & 2.5  & 3    \\ \midrule
$\xi_{1}=-0.1$& 0.2 (0.01) & 0.4 (0.02) & 0.8 (0.05) & 1.2 (0.07) & 1.6 (0.09) & 0.04 (0.01) & 0.1 (0.01) & 0.5 (0.03) & 1.1 (0.07) & 1.8 (0.1) & 0.01 (0.01) & 0.01 (0.01) & 0.2 (0.02) & 0.8 (0.05) & 2.0 (0.1) \\
\rowcolor{Gray}$\xi_{1}=-0.1$ & 0.2 & 0.4 & 0.8 & 1.3 & 1.6 & 0.1 & 0.2 & 0.5 & 1.2 & 1.9 & 0.01 & 0.1 & 0.3 & 0.9 & 1.9 \\
$\xi_{1}=-0.5$ & 1.4 (0.09) & 2.6 (0.2) & 3.0 (0.2) & 2.9 (0.2) & 2.9 (0.2) & 0.9 (0.05) & 2.6 (0.2) & 3.6 (0.2) & 4.6 (0.3) & 4.0 (0.2) & 0.1 (0.01) & 2.2 (0.1) & 4.6 (0.3) & 5.6 (0.4) & 5.7 (0.3) \\
\rowcolor{Gray}$\xi_{1}=-0.5$  & 1.5 & 2.6 & 3.1 & 3.2 & 2.9 & 1.0 & 2.7 & 4.0 & 4.3 & 4.1 & 0.4 & 2.2 & 4.4 & 5.5 & 5.5 \\
$\xi_{1}=-1$ & 3.7 (0.2) & 4.6 (0.3) & 4.5 (0.3) & 4.4 (0.3) & 3.7 (0.2) & 3.1 (0.2) & 6.3 (0.4) & 6.3 (0.4) & 5.9 (0.4) & 4.9 (0.3) & 2.3 (0.2) & 6.3 (0.4) & 7.2 (0.4) & 7.3 (0.5) & 6.3 (0.4) \\
\rowcolor{Gray}$\xi_{1}=-1$  & 3.6 & 4.7 & 4.7 & 4.2 & 3.6 & 3.5 & 5.8 & 6.4 & 5.9 & 5.1 & 2.4 & 6.1 & 7.8 & 7.8 & 6.9 \\
$\xi_{1}=-5$ & 15.6 (0.9) & 11.8 (0.8) & 9.0 (0.5) & 7.2 (0.4) & 6.1 (0.4) & 22.4 (1) & 16.4 (1) & 15.3 (1) & 11.4 (0.7) & 8.7 (0.5) & 22.7 (1) & 21.6 (1) & 18.7 (1) & 14.6 (1) & 10.5 (0.7) \\
\rowcolor{Gray}$\xi_{1}=-5$ & 15.0 & 12.5 & 10.3 & 8.4 & 6.6 & 20.2 & 17.7 & 14.8 & 12.1 & 9.4 & 23.4 & 22.5 & 19.4 & 16.0 & 12.5 \\
$\xi_{1}=-10$ & 20.8 (1) & 15.7 (1) & 14.5 (0.9) & 10.9 (0.7) & 7.3 (0.5) & 30.9 (2) & 26.4 (2) & 21.0 (1) & 14.4 (0.9) & 10.3 (0.6) & 38.1 (2) & 31.5 (2) & 25.7 (2) & 20.5 (1) & 13.8 (0.9) \\
\rowcolor{Gray}$\xi_{1}=-10$  & 23.1 & 17.9 & 14.4 & 11.5 & 8.9 & 32.5 & 25.8 & 20.9 & 16.6 & 12.8 & 40.4 & 33.6 & 27.5 & 22.0 & 16.9 \\
 \bottomrule
\end{tabular}
}
\caption{Simulated relative efficiency of $\hR_{\rm PC}^{(\rm full)}$
with $\pi_1=\pi_2$ for $n=500, p=1$ (Bootstrap standard errors
are in parentheses). The grey shaded rows give for ease of comparison
the values of the ARE from Table 2}
\label{tab:n=500}
\end{table}
\end{landscape}

\nocite{*}

\bibliographystyle{myagsm}
\bibliography{bibliography}

\appendix
\section*{Appendix}
We consider here under the canonical form (\ref{eq:g2b})
of the model the evaluation of the information matrices in the expression for the 
information matrix $\bI_{\rm PC}^{\rm full)}(\bbeta)$, which has to be carried out
to obtain the ARE of the rule $\hR_{\rm PC}^{(\rm full)}$ based on the full ML
estimator of $\bbeta$ formed from the partially classified sample $\bx_{\rm PC}$.
We also provide more details on the proof of Theorem 2, 
in particular, the extension of Theorem 2 to the case of unequal prior probabilities.

\subsection*{The information matrix $\mat{I}_{\text{CC}}(\vect{\beta})$}
\label{subsec:info_cc}
It is shown in \citep{efron_1975_efficiency} that the matrix $\mat{I}_{\text{CC}}(\vect{\beta})$ has the following structure
\begin{align}
  \mat{I}_{\text{CC}}(\vect{\beta}) &= \begin{pmatrix}
    a_{0} & a_{1} & 0 & 0 &\cdots & 0 \\
    a_{1} & a_{2} & 0 & 0 &\cdots& 0 \\
    0 & {0} & a_{3} & {0} &\cdots & {0} \\
    \vdots & \vdots & {0}& \ddots &\cdots  & {0} \\
        {0} & {0} &{0} & {0} &a_{3}  &{0} \\
        {0} & {0}  & {0} & {0}& {0} & a_{3} \\
    \end{pmatrix}, \label{eq:info_cc}
\end{align}
where 
\begin{align}
    \begin{pmatrix}
        a_{0} & a_{1} \\
        a_{1} & a_{2}
    \end{pmatrix}^{-1}
    &=  \dfrac{1}{\pi_{1}\pi_{2}}\begin{pmatrix}
        1+\Delta^2/4 & -(\pi_{2}-\pi_{1})\Delta/2 \\
         -(\pi_{2}-\pi_{1})\Delta/2 & 1+2\pi_{1}\pi_{2}\Delta^2
    \end{pmatrix}, \label{eq:a_definition}
\end{align}
and $a_{3}=\pi_{1}\pi_{2}(1+\Delta^2\pi_{1}\pi_{2})^{-1}$. If $\pi_{1}=\pi_{2}$, $a_{1}$ is zero and so the matrix $  \mat{I}_{\text{CC}}(\vect{\beta})$ is diagonal, and in addition  $a_{0}=a_{3}=\{4(1+\Delta^2/4)\}^{-1}$.
\subsection*{The information matrix $\mat{I}_{\text{CC}}^{\text{(clr)}}(\vect{\beta})$}
\label{subsec:info_clr}
The conditional distribution of $\vect{Y}$ given that $M=1$ can be expressed as
\begin{align}
    f_{\text{miss}}(\vect{y} \mid M =1) &= \dfrac{q_{1}(y_1)f_{y_1}(y_{1})}{\gamma}\prod_{i=2}^{p}\phi(y_{i}; 0, 1). \label{eq:ycond}
\end{align}
The matrix $\mat{I}_{\text{CC}}^{\text{(clr)}}(\vect{\beta})$ is given by the integral
\begin{align*}
\mat{I}_{\text{CC}}^{\text{(clr)}}(\vect{\beta})  &= \int_{R^{p}}\begin{pmatrix}
    1 \\
    \vect{y}
    \end{pmatrix}\begin{pmatrix}
    1 & \vect{y}^{\T} \end{pmatrix} \tau_{1}(y_{1})\tau_{2}(y_{1}) f_{\text{miss}}(\vect{y} \mid M=1) \ d\vect{y}.
\end{align*}
Using the independence of the variables in $\vect{Y}$ in the conditional distribution 
(\ref{eq:ycond}), the matrix has the structure
\begin{align}
\mat{I}_{\text{CC}}^{\text{(clr)}}(\vect{\beta}) &= \begin{pmatrix}
    d_{0} & d_{1} & {0}& {0}&\cdots &{0}\\
    d_{1} & d_{2} & {0} & {0}&\cdots& {0}\\
   {0} &{0} & d_{0} &{0} &\cdots &{0}\\
    \vdots & \vdots & {0} & \ddots &\cdots  & {0} \\
       {0}& {0} & {0}& {0} &d_{0}  &{0}\\
        {0}& {0}  & {0} & {0} & {0}& d_{0} \\
    \end{pmatrix}, \label{eq:info_clr}
\end{align}
where
\begin{align}
d_{k} &= \int_{-\infty}^{\infty}y_{1}^{k}\tau_{1}(y_{1})\tau_{2}(y_1)q_{1}(y_{1}) \gamma^{-1}f_{y_1}(y_{1})\ dy_{1},  \qquad (k=0,1,2), \label{eq:d_definition}
\end{align}
and the functions $\tau_{1}, \tau_{2}, q_{1}$, and $f_{y_1}$ are as given in Theorem 1. For $\pi_{1}=\pi_{2}$, $d_{1}$ is zero and so the matrix $\mat{IC}_{\text{CC}}^{\rm(clr)}(\vect{\beta})$ is diagonal. 
\subsection*{The information matrix $\mat{I}_{\text{PC}}^{\text{(miss)}}(\vect{\beta})$}
\label{subsec:info_miss}
Using the independence of the variables in $\vect{Y}$,
the matrix $\mat{B}_{22}$ has the following structure
\begin{align}
    \mat{B}_{22} &= \begin{pmatrix}
    b_{0} & b_{1} & {0} &{0} &\cdots &{0} \\
    b_{1} & b_{2} &{0} & {0} &\cdots& {0} \\
    {0} & {0} & b_{0} & {0}&\cdots & {0} \\
    \vdots & \vdots & {0} & \ddots &\cdots  & {0}\\
        {0} & {0} &{0} &{0} &b_{0}  &{0} \\
        {0} & {0}  &{0} & {0} & {0} & b_{0} \\
    \end{pmatrix}. \label{eq:info_b22}
\end{align}
The elements of $\mat{B}_{22}$ are given by
\begin{align*}
    b_{0} &= \int_{-\infty}^{\infty}4\xi_{1}^2(\Delta^2y_{1}^{2}+2\lambda\Delta+\lambda^2)q_{1}(y_{1})(1-q_{1}(y_1))f_{y_1}(y_{1})\ dy_{1}, \\
    b_{1} &= \int_{-\infty}^{\infty}\xi_{1}^2(2\lambda+2\Delta y_{1})q_{1}(y_{1})(1-q_{1}(y_1))f_{y_1}(y_{1})\ dy_{1}, \\
        b_{2} &= \int_{-\infty}^{\infty}\xi_{1}^2(2\lambda+2\Delta y_{1})q_{1}(y_{1})(1-q_{1}(y_1))f_{y_1}(y_{1})\ dy_{1}.
\end{align*}

The matrix $\mat{B}_{23}$ has the following structure
\begin{align}
    \mat{B}_{23} &= \begin{pmatrix}
    r_{0} & r_{1} \\
    r_{2} & r_{3} \\
   {0} & {0}\\
    \vdots & \vdots \\
  {0}& {0}
    \end{pmatrix}. \label{eq:info_b23}
\end{align}
The nonzero elements of $\mat{B}_{23} $ are given by
\begin{align*}
    r_{0} &= \int_{-\infty}^{\infty}\xi_{1}(2\lambda+2\Delta y_{1})q_{1}(y_{1})(1-q_{1}(y_{1})) f_{y_1}(y_{1}) \ dy_{1},\\
     r_{1} &= \int_{-\infty}^{\infty}\xi_{1}(\lambda+\Delta y_{1})^2(2\lambda+2\Delta y_{1})q_{1}(y_{1})(1-q_{1}(y_{1}))f_{y_1}(y_{1}) \ dy_{1},\\
     r_{2} &= \int_{-\infty}^{\infty}\xi_{1}(2\lambda y_{1}+2\Delta y_{1}^2) q_{1}(y_{1})(1-q_{1}(y_{1}))f_{y_1}(y_{1})\ dy_{1},\\
          r_{3} &= \int_{-\infty}^{\infty}\xi_{1}(\lambda+\Delta y_{1})^2(2\lambda y_{1}+2\Delta y_{1}^2)q_{1}(y_{1})(1-q_{1}(y_{1}))f_{y_1}(y_{1})\ dy_{1}.
\end{align*}
For $\pi_{1}=\pi_{2}$, $r_{0}$ and $r_{1}$ are both equal to zero as they are equal to the integral of an odd function over the real line. The information matrix for the estimation of $\vect{\xi}$, $\mat{B}_{33}$, can be written as
\begin{align}
    \mat{B}_{33} &= \begin{pmatrix}
    s_{0} &  s_{1} \\
    s_{1} & s_{2}
    \end{pmatrix}, \label{eq:info_b33}
\end{align}
where
\begin{align*}
    s_{0} &= \int_{-\infty}^{\infty}q_{1}(y_{1})(1-q_{1}(y_{1}))f_{y_1}(y_{1}) \ dy_{1}, \\
    s_{1} &= \int_{-\infty}^{\infty}(\lambda+\Delta y_{1})^2q_{1}(y_{1})(1-q_{1}(y_{1})) f_{y_1}(y_{1})\ dy_{1},\\
    s_{2} &= \int_{-\infty}^{\infty}(\lambda+\Delta y_{1})^4q_{1}(y_{1})(1-q_{1}(y_{1})) f_{y_1}(y_{1})\ dy_{1}.\\
\end{align*}
As the lower block of $\mat{B}_{23}$ given by (\ref{eq:info_b23})
is the zero matrix for all $\pi_{1}$, only the top left two-by-two block of $\mat{B}_{23}\mat{B}_{33}^{-1}\mat{B}_{23}$ will be nonzero. Let
\begin{align}
    \begin{pmatrix}
    w_{0} & w_{1} \\
    w_{1} & w_{2}
    \end{pmatrix} &= \begin{pmatrix}
    r_{0} & r_{1} \\
    r_{2} & r_{3} \\
    \end{pmatrix}
    \begin{pmatrix}
    s_{0} & s_{1} \\
    s_{1} & s_{2}
    \end{pmatrix}^{-1}
    \begin{pmatrix}
    r_{0} & r_{1} \\
    r_{2} & r_{3} \\
    \end{pmatrix}^{\T}.
    \label{eq:w_definition}
\end{align}
In general,
\begin{align}
    \mat{I}_{\text{PC}}^{\rm(miss)}(\vect{\beta}) &= \begin{pmatrix}
    b_{0}-w_{0} & b_{1}-w_{1} & {0} &  {0} &\cdots &{0} \\
    b_{1}-w_{1} & b_{2}-w_{2} & {0} &  {0} &\cdots& {0} \\
    {0}& {0}& b_{0} & {0}&\cdots & {0} \\
    \vdots & \vdots & {0} & \ddots &\cdots  &{0} \\
        {0}& {0} & {0} & {0} &b_{0}  &{0} \\
       {0} & {0} & {0} & {0} & {0} & b_{0} \\
    \end{pmatrix}.
\end{align}
 For $\pi_1=\pi_2$, $h_0$ and $h_1$ are both zero, and so then $w_0$ and $w_1$ are also both zero, leading then to the matrix $\mat{I}_{\text{PC}}^{\rm(miss)}(\vect{\beta})$ being diagonal. 
\subsection*{Asymptotic covariance matrix of $\widehat{\vect{\beta}}_{\rm PC}^{\rm{(full)}}$}
Let 
\begin{align}
\begin{pmatrix}
    h_{0} & h_{1} \\
    h_{1} & h_{2}
\end{pmatrix} &= 
\begin{pmatrix}
    a_{0} & a_{1} \\
    a_{1} & a_{2}
\end{pmatrix}-
\gamma(\vect{\Psi})
\begin{pmatrix}
    d_{0} &d_{1} \\
    d_{1} & d_{2}
\end{pmatrix}+
\begin{pmatrix}
    b_{0} & b_{1} \\
    b_{1} & b_{2}
\end{pmatrix}
-
\begin{pmatrix}
    w_{0} &  w_{1} \\
    w_{1} & w_{2}
\end{pmatrix}, \\
u_{0}&=a_{3}-\gamma(\vect{\Psi}) d_{0}+b_{0},
\end{align}
where the constants $a_{0}, a_{1}, a_{2},$ and $a_{3}$ are given in 
(\ref{eq:a_definition}), 
the constants $d_{0}, d_{1}$ and $d_{2}$ are given in 
(\ref{eq:d_definition}), $b_{0}, b_{1}, b_{2}$ are given in 
(\ref{eq:info_b22}) and  $w_{0}, w_{1}, w_{2}$  are given in 
(\ref{eq:w_definition}). The general form of the information matrix is
\begin{align}
    \mat{I}_{\rm PC}^{\rm (full)}(\vect{\beta})&= \begin{pmatrix}
    h_{0} & h_{1} & {0} & {0} &\cdots & {0} \\
    h_{1} & h_{2} & {0} &  {0} &\cdots&  {0} \\
    {0} & {0} & u_{0} & {0} &\cdots & {0}\\
    \vdots & \vdots &{0} & \ddots &\cdots  & \\
      {0}& {0} & {0} & {0} &u_{0}  &{0} \\
        {0} &{0}  & {0} & {0} &  {0} & u_{0} \\
    \end{pmatrix}. \label{eq:info_genera}
\end{align}
As mentioned earlier, for $\pi_{1}=\pi_{2}$  a number of useful simplifications can be made. The matrices $\mat{I}_{\text{CC}}(\bbeta)$, $\bI_{\rm CC}^{(\rm clr)}(\bbeta)$, and $
\bI_{\rm PC}^{(\rm miss)}(\bbeta)$ are diagonal, and  $r_{0}=u_{0}$ and $r_{1}=0$. For $\pi_{1}=\pi_{2}$ the information matrix reduces to
\begin{align}
    \mat{I}_{\rm PC}^{\rm (full)}(\vect{\beta})&= \begin{pmatrix}
    u_{0} &{0}& {0} &{0}&\cdots & {0}\\
    {0}& a_{2}-\gamma(\vect{\Psi}) d_{2}+b_{2}-w_{2} & {0}& {0}&\cdots& {0} \\
    {0}& {0} & u_{0} & {0} &\cdots & {0} \\
    \vdots & \vdots & {0} & \ddots &\cdots  & {0} \\
        {0}& {0}& {0} & {0}& u_{0}  &{0}\\
       {0} & {0}  & {0} & {0}& {0}& u_{0} \\
    \end{pmatrix}. \label{eq:info_equal}
\end{align}
The asymptotic covariance matrix is given by 
$\mat{V}$ is given by $n\{  \mat{I}_{\rm PC}^{\rm (full)}(\vect{\beta})\}^{-1}$. For $\pi_{1}=\pi_{2}$, $v_{jj}=1/u_{0}$ for $j=0,2,3, \ldots, p$.

\subsection*{Extension of Theorem 2 to Arbitrary Prior Probabilities}
\label{subsec:general_are}
We refer to the result (\ref{eq:g37})
given by \cite{efron_1975_efficiency}
for the first order expansion of the expected excess error rate
of the plug-in form of the Bayes' rule using the estimator $\hbbeta$ of $\bbeta$ where
$\sqrt{n}(\hbbeta-\bbeta)$ converges in distribution 
to the $N(\bzero,\bV)$ distributon, 
as $n \rightarrow \infty$ and where 
the first and second order moments also converge.

The expectation of the so-called excess error rate 
can be expanded as 
\begin{eqnarray}
E\{{\rm err}(\hbbeta)\}-{\rm err}(\bbeta)
&=&\frac{\pi_1\phi(\Delta^*;0,1)}{2\Delta n}\,w + o(1/n),
\label{eq:g37a}
\end{eqnarray}
where
$$w=v_{00}-\frac{2\lambda}{\Delta} v_{01}
+\frac{\lambda^2}{\Delta^2}v_{11}
+\sum_{i=2}^p  v_{ii} $$
and where $\lambda=\log(\pi_1/\pi_2), 
\Delta^*={\textstyle\frac{1}{2}}\Delta-\lambda/\Delta$, and 
$\phi(y;\mu,\sigma^2)$ denotes 
the normal density with mean $\mu$ and variance $\sigma^2$.
Here $v_{jk}=(\bV)_{jk}$, where
the columns and rows in $\bV$ are indexed from zero to $p$.

Let
\begin{align*}
    Q_{1} &= \dfrac{1}{\pi_{1}\pi_{2}}\begin{pmatrix}
    1 & -\lambda/\Delta
    \end{pmatrix} \begin{pmatrix}
        1+\Delta^2/4 & -(\pi_{2}-\pi_{1})\Delta/2 \\
         -(\pi_{2}-\pi_{1})\Delta/2 & 1+2\pi_{1}\pi_{2}\Delta^2
    \end{pmatrix}
    \begin{pmatrix}
     1 & -\lambda/\Delta
    \end{pmatrix}^{\T}, \\
    Q_{2} &= \dfrac{1}{\pi_{1}\pi_{2}}(1+\pi_{1}\pi_{2}\Delta^2).
\end{align*}
Using the expansion of the error rate in 
(\ref{eq:g37}), the first order approximation to the expected error rate of $\widehat{R}_{\rm CC}$ is
\begin{align}
     E\{{\rm err}(\hbbeta_{\rm CC})\}
-{\rm err}(\bbeta)  &=\dfrac{\pi_{1}\phi(\Delta^*; 0, 1)}{2\Delta n}\{ Q_{1}+(p-1)Q_{2} \} + o(1/n). \label{eq:ecc}
\end{align}
Let
\begin{align}
\mat{H} &= 
\begin{pmatrix}
    a_{0} & a_{1} \\
    a_{1} & a_{2}
\end{pmatrix}-
\gamma(\vect{\Psi})
\begin{pmatrix}
    d_{0} &d_{1} \\
    d_{1} & d_{2}
\end{pmatrix}+
\begin{pmatrix}
    b_{0} & b_{1} \\
    b_{1} & b_{2}
\end{pmatrix}
-
\begin{pmatrix}
    w_{0} &  w_{1} \\
    w_{1} & w_{2}
\end{pmatrix}, \label{eq:Hmatrix} \\
u_{0} &= a_{3}-\gamma(\vect{\Psi}) d_{0}+b_{0},
\end{align}
where the constants $a_{0}, a_{1}, a_{2},$ and $a_{3}$ are given in 
(\ref{eq:a_definition}), the constants $d_{0}, d_{1}$, and $d_{2}$ 
are given in (\ref{eq:d_definition}), 
$b_{0}, b_{1}, b_{2}$ are given in 
(\ref{eq:info_b22}) and  $w_{0}, w_{1}, w_{2}$  are given 
in (\ref{eq:w_definition}). Define
\begin{align*}
    Q_{3} &= \begin{pmatrix}
    1 & -\lambda/\Delta
    \end{pmatrix} \mat{H}^{-1}\begin{pmatrix}
    1 & -\lambda/\Delta
    \end{pmatrix}^{\T},\\
    Q_{4} &= 1/u_{0}.
\end{align*}
Using the expansion (\ref{eq:g37}), 
the first order approximation to the expected error rate of 
$\widehat{R}_{\rm PC}^{\rm (full)}$ is
\begin{align}
   E\{{\rm err}(\hbbeta_{\rm PC}^{(\rm full)})\}
-{\rm err}(\bbeta) &=\dfrac{\pi_{1}\phi(\Delta^{*}; 0,1)}{2\Delta n}\{ Q_{3}+(p-1)Q_{4} \} + o(1/n),  \label{eq:epc}
\end{align}
which gives the denominator for arbitrary $\pi_1$ 
in the formula (\ref{eq:g34}) for the ARE.
Evaluation of $Q_{3}$ involves some effort, 
as we need to determine each of the constants appearing in the matrix $\mat{H}$ in 
(\ref{eq:Hmatrix}).

Taking the ratio of (\ref{eq:ecc}) to (\ref{eq:epc})
and ignoring terms of $o(1/n)$ gives the asymptotic relative efficiency of 
$\widehat{R}_{\rm PC}^{\rm (full)}$ to  $\widehat{R}_{\rm CC}$,
\begin{align}
    {\rm{ARE}}(\widehat{R}_{\rm PC}^{\rm (full)}) &= \dfrac{Q_{1}+(p-1)Q_{2}}{Q_{3}+(p-1)Q_{4}}. \label{eq:are_general}
\end{align}
Evaluation of \eqref{eq:are_general} involves many calculations due to the number of terms in $Q_{3}$. 
The general form \eqref{eq:are_general} simplifies if $\pi_{1}=\pi_{2}$,  as then
$Q_{1}=Q_{2}= 4(1+\Delta^2/4)$ and $
Q_{3}=Q_{4}=1/u_{0}$. The asymptotic relative efficiency when $\pi_{1}=\pi_{2}$ then collapses to the more interpretable form,
\begin{align}
    {\rm{ARE}}(\widehat{R}_{\rm PC}^{\rm (full)})&= \dfrac{pQ_{2}}{pQ_{4}} \nonumber \\
  &= 4(1+\Delta^2/4)u_{0}, \label{eq:are_final}
\end{align}
which holds for all $p$.

\end{document}